\documentclass[showpacs,prl,onecolumn,aps,superscriptaddress,preprintnumbers,nofootinbib]{revtex4}
\usepackage[T1]{fontenc}
\usepackage[latin9]{inputenc}
\usepackage{graphicx,hyperref}

\usepackage{amsmath,amssymb}
\usepackage{epsfig}
\usepackage{graphicx}
\usepackage{amsmath}
\usepackage{amsfonts}
\usepackage{mathrsfs}
\usepackage{pifont}
\usepackage{relsize}
\usepackage{mathtools}
\def\slashchar#1{\setbox0=\hbox{$#1$}     		
   \dimen0=\wd0                                 	
   \setbox1=\hbox{/} \dimen1=\wd1               	
   \ifdim\dimen0>\dimen1                        	
      \rlap{\hbox to \dimen0{\hfil/\hfil}}      	
      #1                                        	
   \else                                        	
      \rlap{\hbox to \dimen1{\hfil$#1$\hfil}}   	
      /                                         	
   \fi}

\renewcommand{\vec}{\boldsymbol}
\newcommand{\beq}{\begin{equation}}
\newcommand{\eeq}{\end{equation}}
\newcommand{\bea}{\begin{eqnarray}}
\newcommand{\eea}{\end{eqnarray}}
\newcommand{\baa}{\begin{array}}
\newcommand{\eaa}{\end{array}}

\def\eq#1{{Eq.~(\ref{#1})}}

\def\fig#1{{Fig.~\ref{#1}}}
\newcommand{\intl}{\int\limits}

\newcommand{\bas}{\bar{\alpha}_S}

\newcommand{\nn}{\nonumber}

\newcommand{\bg}{ \bar{\gamma}}

\newcommand{\h}{\frac{1}{2}}

\newcommand{\Lb}{\left(}
\newcommand{\Rb}{\right)}
\def\pom{{I\!\!P}}

\renewcommand{\vec}[1]{\boldsymbol{#1}}

\numberwithin{equation}{section}

\begin{document}
\title{ Multiplicity dependence of quarkonia production at ultra high energies}
 \author{E.~ Levin}
\email{leving@tauex.tau.ac.il}
\affiliation{Department of Particle Physics, School of Physics and Astronomy,
Raymond and Beverly Sackler
 Faculty of Exact Science, Tel Aviv University, Tel Aviv, 69978, Israel}

\date{\today}


\begin{abstract}

In this paper we   propose an approach in high energy QCD, which allows us to calculate the inclusive quarkonia production  at ultra high energies. This approach is based on $t$-channel unitarity and on the expressions for dipole densities from the procedure of summing  of large Pomeron loops which we have developed in our previous papers. In the  framework of this approach we discuss  the  
dependence of quarkonia
 production on the multiplicity of the accompanying hadrons. In this paper we used  
 the three gluons fusion mechanism for quarkonia production, without  assuming the multiplicity
 dependence of the saturation scale. We found  the multiplicity distribution of produced gluon with ( $ \mathcal{P}^{J/\Psi}_n$) and without ($\mathcal{P}^{gluon}_n$) the quarkonia production. It turns out that $    \frac{\mathcal{P}^{J/\Psi}_n}{\mathcal{P}^{gluon}_n} \propto  n^2 $ as has been expected and discussed in corresponding publications.

  \end{abstract}
\maketitle

\vspace{-0.5cm}
\tableofcontents

\section{Introduction}
In this paper we continue our  study of three gluon fusion mechanism for $J/\Psi$ production\cite{KMRS,MOSA,LESI,LSS,GOLEPSI}. For the interaction with
 nuclei this mechanism is dominant \cite{KHTU,KLNT,DKLMT,KLTPSI,KMV}; and
it has been demonstrated
 in Ref.\cite{LESI},  that this mechanism gives a
 substantial contribution in hadron-hadron collisions.

\begin{figure}[ht]
     \begin{center}
     \includegraphics[width=  16cm,height=3.7cm]{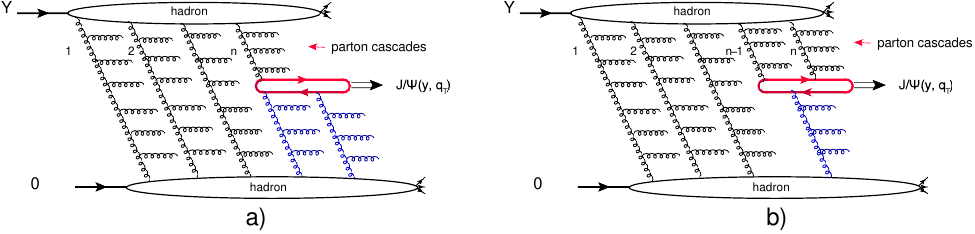} 
     \end{center}  
   \caption{ The production of quarkonia from $n$ parton cascades.}
\label{nlad}
   \end{figure}

The paper has three goals. The first is to develop the approach in high energy QCD to calculate the inclusive cross section using the t-channel unitarity constraints and the expression for the dipole densities. The second
 is to
take into account the general structure of the parton cascade. In particular it accounts the contribution of  production of many Pomeron ladders in the parton cascade (see \fig{nlad}) which cancel each other in the inclusive production of $J/\Psi$ . For example, production of gluons  from extra Pomeron ladders in \fig{nlad}-a  does not contribute to the inclusive production which can be estimated used the diagram of \fig{3p}-b.  Such contributions have been neglected  in all papers on the subject (see for example Res.\cite{LSS,GOLEPSI}). The third goal is to find the multiplicity distribution for $J/\Psi$ production at ultra high energies in  the framework of high
 energy QCD 
 (see Ref.\cite{KOLEB} for a general review). Effective theory of high energy QCD exists
in two different formulations:
   the CGC/saturation \footnote{CGC stands for colour glass condensate.} approach
 \cite{MV,MUCD,B,K,JIMWLK,GIJMV}, and the BFKLL\footnote{BFKL stands for Balitsky,Fadin,Kuraev and Lipatov.} Pomeron calculus 
\cite{BFKL,LI,GLR,GLR1,MUQI,MUPA,BART,BRN,KOLE,LELU1,LELU2,LMP,
AKLL,AKLL1,LEPP}. 
 In 
  this paper we restrict ourself   to  the BFKL 
Pomeron
 calculus, which allows us to apply the AGK\footnote{AGK stands for Abramovsky,Gribov and Kancheli cutting rules.} \cite{AGK} cutting rules for finding of the multiplicity of the produced gluons in the final state.
 Unfortunately, as far as we know, these observables 
 are out of the reach   for  the CGC 
approach. 

In our recent papers\cite{LEDIDI,LEDIA,LEAA} we summed the large Pomeron loops at ultra high energy using the MPSI\cite{MPSI}\footnote{MPSI stands for Mueller, Patel, Salam and Iancu approach.} approach. It turns out that all three types of scattering: dilute-dilute dipoles systems ( dipole-dipoles scattering), dilute-dense dipole system ( dipole-nucleus scattering )    and dense-dense dipole system (ion-ion scattering ), show the same energy dependence due to contributions of  large BFKL Pomeron loops. In this paper we concentrate our effort on the multiplicity distribution of all these processes at ultra high energies.
  
 The recent experiments by  ALICE\cite{ALICE0,ALICE1,ALICE01,ALICE2,ALICE3} 
 and STAR\cite{STAR1,STAR2}, have shown  that the cross sections for $J/\Psi$
 production   strongly depends  on the multiplicity of accompanying 
hadrons.
 These data have  stimulated theoretical discussions on the origin of such
 dependence (see Refs.\cite{KPPRS,FEPA,MTVW,LESI,LSS,GOLEPSI}). 

In the next section of this paper we will discuss the structure of the BFK parton cascade in the simple model for the BFKL pomeron calculus: the BFKL Pomeron calculus in zero transverse  dimension.
Zero transverse dimension toy models can be viewed as a realization of the Pomeron calculus or more generally of 
Reggeon Field Theory (RFT). Over the years they have been intensively used to model high energy collisions in QCD. 
These models\cite{ACJ,AAJ,JEN,ABMC,CLR,CIAF,MUSA,nestor,RS,SHXI,KOLEV,BIT,LEPRI,utm,utmm,utmp} encode various fundamental features of QCD such as unitarity, but are much simpler than the latter and frequently 
solvable analytically.  Hence they provide a valuable playground to gain intuition about the dynamics of real QCD. We believe, that  by using these models, we clarify all aspects of the multiplicity distribution for $J/\Psi$ production.

 In the rest of the paper we discuss the calculation of the inclusive production in the framework of high energy QCD. Our approach is based on the $t$-channel unitarity and on the expression for the dipole densities which we suggested in our previous papers\cite{LEDIDI,LEDIA,LEAA}. We wish to emphasize that this is the first calculation in the framework of such approach.  The advantage of it is in the clear understanding of the typical scales that govern this process. In this paper we restrict ourself by calculating the cross sections integrated over transverse momenta.
 
   In section III we review this approach for the scattering amplitude. In sections IV and V we  propose how to apply the $t$-channel unitarity to calculation of the inclusive production. We calculate the multiplicity distribution for the produced gluons using the AGK \cite{AGK} cutting rules. In section VI we develop the approach for calculation of the multiplicity distribution for $J/\Psi$ production  in three gluon fusion mechanism. In the conclusions we resume our results.

\section{Multiplicity distribution for $J/\Psi$ production in the BFKL cascade in zero transverse dimension.}
In the parton model\cite{FEYN,BJP,Gribov} all partons have average transverse
 momentum 
which does not depend
 on energy. Therefore, we can obtain the parton model from the QCD
 cascade assuming that the unknown confinement of gluons leads to 
the QCD cascade for a dipole of fixed size. In this case the QCD 
 cascade equation (see equations in Ref.\cite{LELU2})  takes the following simple form:
\beq \label{EQPM}
 \frac{d P_n\Lb Y\Rb}{d Y}\,\,=\,\,- \Delta\,n\,  P_n\Lb Y\Rb \,\,+\,\,\Lb n - 1 \Rb \Delta P_{n-1}\Lb Y .\Rb
\eeq
where  $P_n\Lb Y \Rb$ denotes the probability to find $n$ dipoles
 (of a fixed size in our model) at  rapidity $Y$, and  $\Delta$ is
 the intercept of the BFKL Pomeron. 

We can introduce the generating function:

\beq \label{ZF}
Z\Lb Y, u\Rb\,\,=\,\,\sum_n \,P_n\Lb Y\Rb \,u^n 
\eeq
where $u$ are  arbitrary numbers. Using this generating function  we can rewrite  \eq{EQPM} in the  following form for the generating function:
\beq \label{TM4}
\frac{\partial  Z\Lb Y, u\Rb}{ \partial Y}\,\,=\,\,- \Delta \, u \Lb 1 - u\Rb \frac{\partial Z\Lb Y, u\Rb}{\partial u} .
\eeq

At the initial rapidity $Y=0$, we have only one dipole,  so $P_1\Lb Y = 0
 \Rb =1$ and $P_{n > 1} \,=\,0$ (so the state is  the only one dipole);
  at $u = 1$,  $Z\Lb  Y, u=1\Rb\,\,=\,\,\sum_n P\Lb y \Rb\,=\,1$. These
 two properties  determine the initial and the boundary conditions for
 the generating function \beq \label{TM3}
Z\Lb Y = 0,u\Rb \,\,=\,\,u;~~~~~~~~~Z\Lb Y, u=1\Rb\,\,=\,\,1.
\eeq
 

 The general solution to \eq{TM4} is an arbitrary function 
 ($  Z\Lb z \Rb$) of the new variable: $z\,\,=\,\,\Delta\,Y\,\,+\,\,f(u)$, 
with
  f(u)  from the following equation:
\beq \label{TM5}
1\,\,=\,\,-\,u\,\Lb 1\,-\,u\Rb\,f'_u\Lb u\Rb  ~~~~~f\Lb u\Rb\,\,=\,\,\ln\Lb \frac{u\,-\,1}{u}\Rb\,\,+\,\,C_1
\eeq
The form of arbitrary function stems from the initial condition of \eq{TM3} 
\beq \label{TM6}
Z\Lb z\Lb Y=0\Rb\Rb\,=\,u;  
\eeq
Since $u\,\,=\,\,1\Big{/}\Lb 1\,-\,e^z\Rb$ we obtain that
\beq \label{SOLTM}
Z_1\Lb Y,\,u\Rb\,\,=\,\,\frac{u\,\,e^{\,-\,\Delta\, Y}}{1\,\,+\,\,u\,\,
\Lb e^{\,-\,\Delta\,Y}\,-\,1\Rb}\,\,=\,\,\,e^{\,-\,\Delta\, Y}\,\sum^\infty_{n=1}\,u^n \Lb 1\,-\, e^{\,-\,\Delta\,Y}\Rb^{n \,-\,1} .
\eeq
Note, that $Z\Lb Y,u=1\Rb = 1$, as it should be from \eq{TM3}. Recall, that $Z_1$  describes the dipole cascade of the fast dipole.

From \eq{SOLTM} we obtain
\beq \label{SOLTM1}
P_n\Lb Y\Rb\,\,=\,\,e^{ - \Delta\,Y} \Lb 1\,\,-\,\,e^{ - \Delta\,Y}\Rb^{n-1}\,=\,\frac{1}{N\Lb Y\Rb}  \Lb 1\,\,-\,\,\,\frac{1}{N\Lb Y\Rb}\Rb^{n-1}\eeq
at rapidity $Y$. In \eq{SOLTM1} $N\Lb Y\Rb\,=\exp\Lb \Delta\,Y\Rb$ is the average number of dipoles at rapidity $Y$.

     \begin{figure}[ht]
     \begin{center}
     \includegraphics[width=0.6\textwidth]{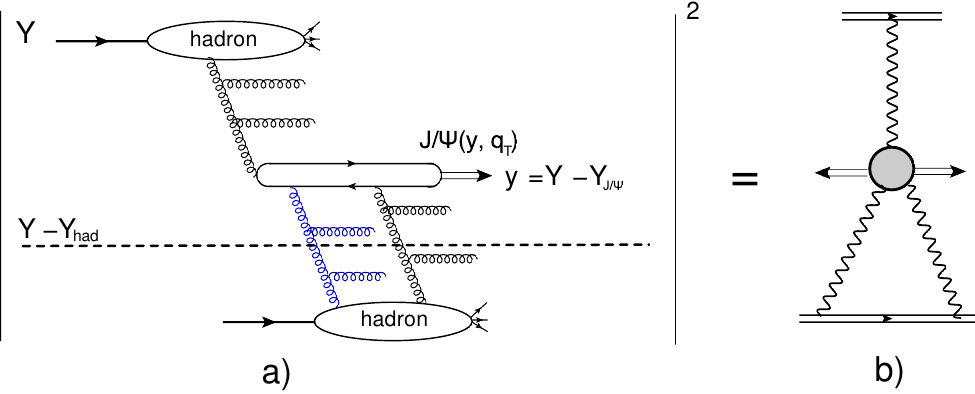} 
     \end{center}    
      \caption{ \fig{3p}-a: The three gluon fusion mechanism
 of $J/\Psi$ production.\fig{3p}-b: the Mueller diagram\cite{MUDIA}
 for $J/\Psi$ production, which illustrates the inter-relation between
 three gluon fusion and the triple BFKL Pomeron interaction. The wavy 
lines
 describe the BFKL Pomerons, while the helical curves represent 
gluons.}
\label{3p}
   \end{figure}
Hence from 
\eq{SOLTM1} the multiplicity distribution of dipoles  at rapidity   $Y_{J//\Psi}$ at which $J/\Psi$ is produced, is equal to:
\beq \label{SOLTM1}
P_m\Lb Y_{J/\Psi}\Rb\,\,=\,\,\frac{1}{N\Lb Y_{J/\Psi}\Rb}  \Lb 1\,\,-\,\,\,\frac{1}{N\Lb Y_{J/\Psi}\Rb}\Rb^{m-1}\eeq
Each of this $m$ dipoles produces its own cascade which can be found from \eq{TM4} with the initial condition:
\beq \label{ICTM1}
Z\Lb Y = Y_{J/\Psi}, u  \Rb= u^m\frac{1}{N\Lb Y_{J/\Psi}\Rb}  \Lb 1\,\,-\,\,\,\frac{1}{N\Lb Y_{J/\Psi}\Rb}\Rb^{m-1}
\eeq
The solution for such initial condition has been found in Ref.\cite{KLL1} and it takes the form for dipoles at rapidity $Y_{had}$\footnote{$Y_{had}$ is the rapidity of hadrons that are measuared.}:
\bea \label{SOLTM2}
&&Z_m\Lb Y = Y_{had} - Y_{J/\Psi}, u \Rb=\sum_{n=m}^\infty u^n P^m_n\Lb \delta
Y = Y_{had} - Y_{J/\Psi},Y_{J/\Psi} \Rb\,\,=\,\,\frac{1}{N\Lb Y_{J/\Psi}\Rb}  \Lb 1\,\,-\,\,\frac{1}{N\Lb Y_{J/\Psi}\Rb}\Rb^{m-1}\Bigg( Z_1\Lb \delta Y\Rb\Bigg)^{m}\nn\\
&&\mbox{with}~ 
P^m_n\Lb \delta Y,Y_{J/\Psi}\Rb\,\,=\,\,\frac{1}{N\Lb Y_{J/\Psi}\Rb}  \Lb 1\,\,-\,\,\frac{1}{N\Lb Y_{J/\Psi}\Rb}\Rb^{m-1}\,\frac{(n-1)!}{(m-1)! \,(n-m )!}\Lb \frac{1}{N\Lb \delta Y\Rb}\Rb^m \Lb 1 - \frac{1}{N\Lb \delta Y\Rb}\Rb^{n - m}
\eea

It is easy to see that
\beq \label{SOLTM3}
\sum_{m=1}^n P^m_n\Lb \delta Y,Y_{J/\Psi}\Rb\,\,=\,\,\frac{1}{N\Lb Y_{had}\Rb}  \Lb 1\,\,-\,\,\,\frac{1}{N\Lb Y_{had}\Rb}\Rb^{n-1}\eeq
In other words, summing over $m$ we obtain  $Z_1\Lb Y_{had}\Rb$ for dipole cascade.

For inclusive production of $J/\Psi$ we need to take into account:( i) that $J/\Psi$ can be produced by any of $m$ dipoles at $ Y - Y_{J/\Psi}$; and (ii) the three gluons fusion  mechanism of $J/\Psi$ production \cite{LESI,LSS,GOLEPSI} (see \fig{3p}). The generating function, which includes the three gluons fusion  mechanism, has the form:
\bea \label{SOLTM4}
Z_{J/\Psi}\Lb \delta Y, Y_{J/\Psi}, u \Rb\,\,&=&\,\,\sum_{m=1}^{\infty}\,\,P_m\Lb Y_{J/\Psi}\Rb\underbrace{\Bigg(Z_1\Lb \delta Y, u\Rb\Bigg)^{m+1 } }_{\mbox{three \,\,gluons \,\,fusion}}\,\,\nn\\
&=&\,\,\frac{\gamma(1 - \gamma) ^2}{\Lb 1 + \gamma \Lb N\Lb \delta Y\Rb\,-\,1\Rb\Rb\,\Lb1 + \gamma \Lb N\Lb  Y\Rb\,-\,1\Rb\Rb} \,=\,\,Z_1\Lb N\Lb Y , u\Rb \Rb\,Z_1\Lb N\Lb \delta Y , u\Rb \Rb \eea
where $\gamma = 1 - u$.

For finding the multiplicity distribution of produced dipoles we need to apply to \eq{SOLTM4} the AGK cutting rules\cite{AGK}. For $Z_1$ it has been done in Refs.\cite{LEPRY,LEPRODTM}. It has been shown in these references that $Z_1$ generates the following inelastic cross sections for production of $n$ cut Pomerons with average multiplicity $n_\pom = n\,\Delta\,Y$:

\beq \label{SOLTM5}
\sigma_n\Lb N\Lb Y\Rb\Rb\,\,=\,\, \frac{\Lb 2 \,\gamma\,N\Lb Y \Rb\Rb^n}{\Lb 1\,+\,2\,\gamma\Lb N\Lb Y\Rb\,-\,1\Rb\Rb^{n-1}}
\eeq

Using \eq{SOLTM5} we can write the inelastic cross section for generating function of \eq{SOLTM4}\footnote{In one dimensional model $\sigma_n$ are dimensionless. The natural generalization for QCD is
$\sigma^{J/\Psi}_n\Lb N\Lb Y\Rb,N\Lb \delta  Y\Rb\Rb\,\,=\,\int d^2 b \sum_{k=1}^{k=n-1} \sigma_k\Lb  N\Lb Y \Rb; b\Rb\,\sigma_{n-k}\Lb  N\Lb \delta Y\Rb; b\Rb$ }:

\beq \label{SOLTM6}
\sigma^{J/\Psi}_n\Lb N\Lb Y\Rb,N\Lb \delta  Y\Rb\Rb\,\,=\,\, \sum_{k=1}^{k=n-1} \sigma_k\Lb  N\Lb Y\Rb\Rb\,\sigma_{n-k}\Lb  N\Lb \delta Y\Rb\Rb\eeq

For calculating the inclusive cross section for  $J/\Psi$    production at fixed multiplicity $n$ ( $\frac{ \sigma^{J/\Psi}_n}{d Y_{J/\Psi}}$) by any of $m$ dipoles at $ Y - Y_{J/\Psi}$ we need
\beq \label{SOLTM7}
\frac{ d \sigma^{J/\Psi}_n\Lb N\Lb Y\Rb,N\Lb \delta  Y\Rb\Rb}{d Y_{J/\Psi}}\,\,=\,\,\,\, \sum_{k=1}^{k=n-1} \,k\,\sigma_k\Lb  N\Lb Y\Rb\Rb\,\sigma_{n-k}\Lb  N\Lb \delta Y\Rb\Rb
\eeq

Summing over all possible $n$ we obtain the inclusive cross section for $J/\Psi$ production, which is equal to
\beq \label{SOLTM8}
\frac{ d \sigma^{J/\Psi}_{incl}\Lb N\Lb Y\Rb,N\Lb \delta  Y\Rb\Rb}{d Y_{J/\Psi}}\,\,=\,\,\frac{ 4\,\gamma^2 N\Lb Y \Rb \,N\Lb \delta Y\Rb}{(1- \gamma)^3 \,\Lb 1\,+\,2\gamma\Lb N\Lb \delta Y\Rb\,-\,1\Rb\Rb}
\eeq

 For small $\gamma$ \eq{SOLTM8} leads to $- \gamma^2 N\Lb \delta Y\Rb\,N\Lb Y\Rb$ which corresponds to the diagram of \fig{3p}-b.

 The analytic form of \eq{SOLTM7} is rather cumbersome. We write here the expression at small $\gamma$:
 \bea \label{SOLTM9}
&&\frac{ d \sigma^{J/\Psi}_n\Lb N\Lb Y\Rb,N\Lb \delta  Y\Rb\Rb}{d Y_{J/\Psi}}\,\,=\\
&&\, \frac{ 1}{\Lb N\Lb Y\Rb\,-\,N\Lb \delta Y\Rb\Rb^2}\Bigg\{  N\Lb Y\Rb\,\Lb \Lb \frac{ 2 \,\gamma\,N\Lb \delta Y\Rb}{1+
 2 \,\gamma\,N\Lb \delta Y\Rb}\Rb^n \,-\, \Lb \frac{ 2 \,\gamma\,N\Lb Y\Rb}{1+
 2 \,\gamma\,N\Lb  Y\Rb}\Rb^ n\Rb \,+\,n\,\frac{\Lb  N\Lb Y\Rb\,-\,N\Lb \delta Y\Rb\Rb}{ 1\,+\,2\,\gamma\,N\Lb \delta Y\Rb} \Lb \frac{ 2 \,\gamma\,N\Lb Y\Rb}{1+
 2 \,\gamma\,N\Lb  Y\Rb}\Rb^n \Bigg\}\nn
 \eea
We can illustrate the main features of this equation writing it for $N\Lb  Y\Rb =N\Lb  \delta Y\Rb$\footnote{Actually, this is a measured case when $J/\Psi$ produced in the same rapidity window where the multiplicity distribution of produced dipoles is measured.}:
 \beq \label{SOLTM10}
\frac{ d \sigma^{J/\Psi}_n\Lb N\Lb Y\Rb,N\Lb \delta  Y\Rb\Rb}{d Y_{J/\Psi}}\,\,=\,\,\frac{n ( n - 1)}{2\Lb 1 + 2\,\gamma N\Lb \delta Y\Rb\Rb} \sigma_n\Lb \delta Y\Rb
\eeq
For the ratio $ \frac{ d \sigma^{J/\Psi}_n\Lb N\Lb Y\Rb,N\Lb \delta  Y\Rb\Rb}{d Y_{J/\Psi}}\Bigg{/}\frac{ d \sigma^{J/\Psi}_{incl}\Lb N\Lb Y\Rb,N\Lb \delta  Y\Rb\Rb}{d Y_{J/\Psi}} $ we have
 \beq \label{SOLTM11}
  \frac{ d \sigma^{J/\Psi}_n\Lb N\Lb Y\Rb,N\Lb \delta  Y\Rb\Rb}{d Y_{J/\Psi}}\Bigg{/}\frac{ d \sigma^{J/\Psi}_{incl}\Lb N\Lb Y\Rb,N\Lb \delta  Y\Rb\Rb}{d Y_{J/\Psi}}\,\,= \,\,\,\,\frac{n ( n - 1)}{2\,\Lb 2\,\gamma\,N\Lb Y\Rb\Rb^2} \sigma_n\Lb \delta Y\Rb
  \eeq
  This result  is the same as it was discussed in the previous publications\cite{KMRS,MOSA,LESI,LSS,GOLEPSI}  . Therefore, it indicates that the corrections from the production of multi ladders in \fig{nlad} leads to a small corrections which can be treated perturbatively. 
  
\section{A recap: summing large Pomeron loops in   the saturation region of   dipole-dipole scattering. }
\begin{figure}[ht]
    \centering
  \leavevmode
      \includegraphics[width=0.3\textwidth]{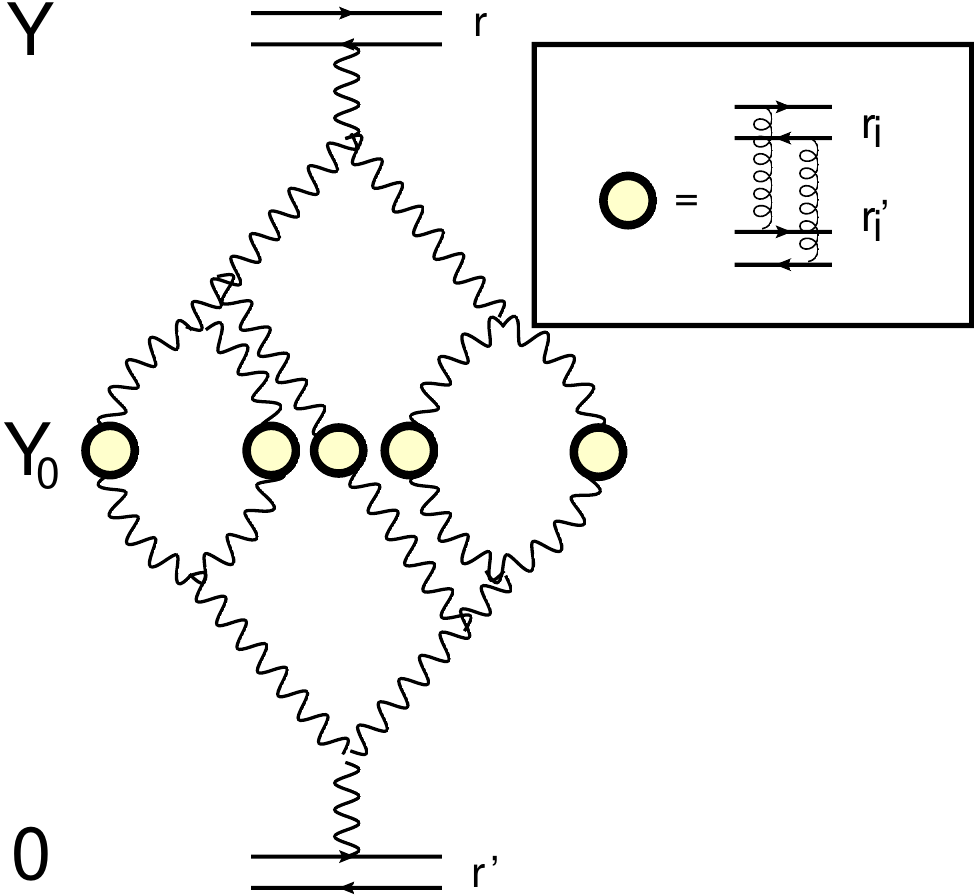}  
      \caption{Summing  large Pomeron loops for dipole-dipole scattering. The wavy lines denote the  BFKL Pomeron exchanges.  The circles denote the amplitude $\gamma$ in the Born approximation of perturbative QCD.  }
\label{mpsi}     
   \end{figure}

In our recent paper\cite{LEDIDI} we have found the contribution to the dipole-dipole scattering amplitude of  the sum of  large Pomeron loops. Here we give a brief review of this approach since we are going to use it  below  for finding the multiplicity distributions of produced gluons. Our approach is based on the $t$-channel unitarity, which  has been written in the convenient for the BFKL Pomeron calculus in Refs.\cite{MPSI,KOLEB,MUDI,LELU1,LELU2,KO1,LE11}(see \fig{mpsi}).
 The analytic expression  for arbitrary $Y_0$ takes the form
       \cite{LELU1,LELU2,KO1,LE11}: 
        \bea \label{MPSI}
     && A\Lb Y, r, r' ;  \vec{b}\Rb\,=\\
     &&\,\sum^\infty_{n=1}\,\Lb -1\Rb^{n+1}\,n!\int  \prod^n_i d^2 r_i\,d^2\,r'_i\,d^2 b'_i 
     \int \!\!d^2 \delta b_i\, \gamma^{BA}\Lb r_i,r'_i, \vec{b}_i -  \vec{b'_i}\equiv \delta \vec{b} _i\Rb 
    \,\,\rho^P_n\Lb Y - Y_0,r,b,  \{ \vec{r}_i,\vec{b}_i\}\Rb\,\rho^T_n\Lb Y_0,r', b', \{ \vec{r}'_i,\vec{b}'_i\}\Rb \nn
      \eea
  $\gamma^{BA}$ is the scattering amplitude of two dipoles in the Born approximation of perturbative QCD.  The dipole  densities: $\rho^P_i \Lb Y -Y_0 , r, b, \{ \vec{r}_i,\vec{b}_i\}\Rb$  for projectile and $\rho^T_i\Lb Y_0 , r', b', \{ \vec{r}'_i,\vec{b}_i\}\Rb$  for target, have been introduced in Ref.\cite{LELU1}  as follows:
\beq \label{PD}
\rho_n( Y\,-\,Y_0, r, b, r_1, b_1\,
\ldots\,,r_n, b_n)\,=\,\frac{1}{n!}\,\prod^n_{i =1}
\,\frac{\delta}{\delta
u_i } \,Z\left(Y\,-\,Y_0;\,[u] \right)|_{u=1}
\eeq
  where  the generating functional $Z$ is
  \beq \label{Z}
Z\Lb Y, \vec{r},\vec{b}; [u_i]\Rb\,\,=\,\,\sum^{\infty}_{n=1}\int P_n\Lb Y,\vec{r},\vec{b};\{\vec{r}_i\,\vec{b}_i\}\Rb \prod^{n}_{i=1} u\Lb \vec{r}_i\,\vec{b}_i\Rb\,d^2 r_i\,d^2 b_i
\eeq
 where $u\Lb \vec{r}_i\,\vec{b}_i\Rb \equiv\,u_i$ is an arbitrary function and $P_n$ is the probability to have $n$ dipoles with the  given kinematics.
 The initial and  boundary conditions for the BFKL cascade  stem from one dipole has 
the following form for the functional $Z$:
\begin{subequations}
\bea
Z\Lb Y=0, \vec{r},\vec{b}; [u_i]\Rb &\,\,=\,\,&u\Lb \vec{r},\vec{b}\Rb;~~~~~~~~Z\Lb Y, r,[u_i=1]\Rb = 1;\label{ZIC}\\
\rho_1\Lb Y=0,   r,b, r_1,b_1\Rb\,\,&=&\,\,\delta^{(2)}\Lb \vec{r} - \vec{r}_1\Rb \delta^{(2)}\Lb \vec{b} - \vec{b}_1\Rb ;~~~~~\rho_n\Lb 
 Y=0, \vec{r},\vec{b}; [r_i, b_i]\Rb \,=\,0 ~ \mbox{at}~\,n\geq 2;\label{ZSR}
\eea
\end{subequations}
 In \eq{MPSI} $\vec{b}_i\,\,=\,\,\vec{b} \,-\,\vec{b'}_i$. 
 Comparing \eq{PD} and \eq{Z} one can see that  $\rho_n = M_n/n!$ where $M_n$ is the factorial moment for the scattering amplitude. Factorial moments are the natural   observables for the BFKL Pomeron calculus (see Ref.\cite{MUDI}).

The main ingredient of \eq{MPSI} is the dipole densities, which have been found in Re.\cite{LEDIDI}\footnote{In Ref.\cite{LEDIDI} \eq{DD5} is written with $\omega$ instead of $\omega - 1$. We will discuss this issue below in section IV}:

\beq \label{DD5}
\rho^P_n\Lb Y - Y_0;  r,b, \{ r_i,b_i\}\Rb = C \,\intl^{\epsilon + i \infty}_{\epsilon - i \infty} \frac{d \omega}{2\,\pi\,i} e^{ \frac{ 
\bg\,\kappa}{2}\,(\omega-1)^2}\frac{1}{n!} \frac{\Gamma\Lb \omega +n\Rb}{\Gamma\Lb \omega \Rb}  \prod^n_{i=1} 
\rho_1\Lb z_i \Rb
\eeq
where
\beq \label{zi}
z_i \,\,=\,\,\,\,\bas \frac{\chi\Lb \bg\Rb}{\bg} \Lb Y \,-\,Y_0\Rb \,\,+\,\,\xi_{r,r_i}: ~~~
\xi_{r,r_i} \,\,=\,\,\ln\Lb \frac{
 r^2\,r_i^2}{\Lb \vec{b}  + \h(\vec{r} - \vec{r}_i)\Rb^2\,\Lb \vec{b} 
 -  \h(\vec{r} - \vec{r}_i)\Rb^2}\Rb\eeq
$\rho_1\Lb z_i\Rb$ satisfies the linear BFKL equation which has a simple solution in the saturation region\cite{GLR,MUT}:
\beq \label{DD1}
N_{\pom}\Lb z_i \Rb  = N_0\,\Lb r^2_i\,Q^2_s\Lb Y - Y_0,b\Rb\Rb^{\bg}\,\,=\,\,N_0 e^{\bg \,z_i}\,\,\equiv \,\,N_0 G_\pom\Lb z_i\Rb\eeq
where $\bg = 1 - \gamma_{cr}$ , $N_0$ is a constant ,  $z_i$ is defined as
\beq \label{zi1}
z_i\,\,=\,\,\ln\Lb r^2_i \,Q^2_s\Lb Y-Y_0,b\Rb\Rb
\eeq
In \eq{DD1} $G_\pom\Lb z_i\Rb$ is the BFKL Pomeron Green's function.

   It can be seen from \eq{zi1}   that 
 BK equation  leads to a new dimensional scale: saturation momentum\cite{GLR}  which has the following $Y$ dependence\cite{GLR,MUT,MUPE}:
 \beq \label{QS}
 Q^2_s\Lb Y, b\Rb\,\,=\,\,Q^2_s\Lb Y=0, b\Rb \,e^{\bas\,\kappa \,Y -\,\,\frac{3}{2\,\gamma_{cr}} \ln Y }
 \eeq 
 where $Y=0$ is the initial value of rapidity and $\kappa$ and $\gamma_{cr}$   are determined by the following equations:
  \beq \label{GACR}
\kappa \,\,\equiv\,\, \frac{\chi\Lb \gamma_{cr}\Rb}{1 - \gamma_{cr}}\,\,=\,\, - \frac{d \chi\Lb \gamma_{cr}\Rb}{d \gamma_{cr}}~
\eeq
where $\chi\Lb \gamma\Rb$ is given by
\beq \label{CHI}
\omega\Lb \bas, \gamma\Rb\,\,=\,\,\bas\,\chi\Lb \gamma \Rb\,\,\,=\,\,\,\bas \Lb 2 \psi\Lb 1\Rb \,-\,\psi\Lb \gamma\Rb\,-\,\psi\Lb 1 - \gamma\Rb\Rb\eeq 
$\psi(z)$ is the Euler $\psi$ function (see Ref.\cite{RY} formulae {\bf 8.36}). As one can see \eq{zi1}  gives the physics meaning of \eq{zi}. In \eq{DD5} 
\beq \label{RHO1}
\rho_1\Lb z_i\Rb = G_\pom\Lb z_i\Rb\eeq

 Coming back to \eq{DD5}, one can see that this equation  
  reconciles the exact solution to the Balitsky-Kovchegov (BK) equation\cite{LETU} , which describes the rare fluctuation in the dipole-target scattering, with the fact that this equation sums the 'fan' Pomeron diagrams.     

The $t$-channel unitarity for the Green's function of the BFKL Pomeron has the form (see Refs.\cite{MUDI,MUSA} 
 \beq \label{TU2}
\hspace{-0.15cm}G_{\pom} \Lb Y, r,r' ;  \vec{b}\Rb =\intl d^2 r_1 d^2 b_1 d^2r'_1 d^2 b'_1
G_\pom\Lb Y - Y_0; r, b -b_1, r_1\Rb \gamma^{BA}\Lb r_1 ,r'_1,\delta b= \vec{b}_1 - \vec{b}'_1\Rb G_\pom\Lb  Y_0; r',0,b'_1, r'_1\Rb
\eeq

Plugging \eq{DD5} into \eq{MPSI} we can calculate the scattering amplitude:

 \bea \label{SA3}
S\Lb z\Rb &=&
C^2\sum^\infty_{n=0} \frac{\Lb - 1\Rb^n}{n!}\!\!\!  \intl^{\epsilon + i \infty}_{\epsilon - i \infty}\!\!\! \frac{ d \omega}{2\,\pi\,i} \!\!\!\intl^{\epsilon + i \infty}_{\epsilon _ i \infty}\!\!\!  \frac{d \omega_1}{2\,\pi\,i} e^{ \frac{ \bg^2 \kappa}{2}\Lb (\omega-1)^2 + (\omega_1-1)^2\Rb}\\
&\times&\frac{ \Gamma\Lb \omega+n\Rb}{\Gamma\Lb \omega\Rb}\frac{ \Gamma\Lb \omega_1+n\Rb}{\Gamma\Lb \omega_1\Rb}\intl\!\!\! d^2 r_i d^2 r'_i d^2b_i d^2b'_i\prod^n_{i=1} 
G_{\pom}\Lb z_i\Rb\gamma^{BA}\Lb r_i,r'_i,\delta b\Rb\,G_{\pom}\Lb z'_i\Rb\nn\eea
 
In   \eq{SA3}  $z_i$ is given in \eq{zi} while $z'_i \,\,=\,\,\,\,\bas \frac{\chi\Lb \bg\Rb}{\bg} \Lb Y_0\Rb \,\,+\,\,\xi_{r',r'_i}$. $\xi_{r',r'_i}$ is equal to $\xi_{r,r_i}$ of \eq{zi} where $r$  and $r_i$ are  replaced by $r'$ and $r'_i$, respectively.

  Using \eq{TU2} we reduce \eq{SA3} to the following expression:
  \beq \label{SA31}
S\Lb z'\Rb =
C^2\sum^\infty_{n=0} \frac{\Lb - 1\Rb^n}{n!}\!\!\!  \intl^{\epsilon + i \infty}_{\epsilon - i \infty}\!\!\! \frac{ d \omega}{2\,\pi\,i} \!\!\!\intl^{\epsilon + i \infty}_{\epsilon _ i \infty}\!\!\!  \frac{d \omega_1}{2\,\pi\,i} e^{ \frac{ \bg^2 \kappa}{2}\Lb( \omega-1)^2 + (\omega_1-1)^2\Rb}\frac{ \Gamma\Lb \omega+n\Rb}{\Gamma\Lb \omega\Rb}\frac{ \Gamma\Lb \omega_1+n\Rb}{\Gamma\Lb \omega_1\Rb}G^n_{\pom}\Lb z'\Rb\eeq
 with 
 \beq \label{SA4}
 z' \,\,= \,\,\,\,\bas \frac{\chi\Lb \bg\Rb}{\bg} \,Y \,\,+\,\,\xi_{r,r'}    
 \eeq   
  $\xi_{r,r'}$ is equal to $\xi_{r,r_1}$ of \eq{zi} where $r$  and $r_1$ are  replaced by $r$ and $r'$, respectively.
   
  Replacing  $\Gamma\Lb \omega_1 +n\Rb$ by the integrals over $t$, viz.:
 \beq \label{DA21}
 \Gamma\Lb \omega_1 +n\Rb\,\,=\,\,\intl^\infty_0 d t\, t^{ \omega_1 +  n -1} e^{- t}
 \eeq  
 we obtain
  \beq \label{SA5}
S\Lb z'\Rb =C^2\,
\sum^\infty_{n=0} \frac{\Lb - 1\Rb^n}{n!}\!\!\!  \intl^{\epsilon + i \infty}_{\epsilon - i \infty}\!\!\! \frac{ d \omega}{2\,\pi\,i} \!\!\!\intl^{\epsilon + i \infty}_{\epsilon -i \infty}\!\!\!  \frac{d \omega_1}{2\,\pi\,i} e^{ \frac{ \bg^2 \kappa}{2}
\Lb (\omega-1)^2 + (\omega_1 -1)^2\Rb}\intl^\infty_0 d t\,  e^{-t}  t^{\omega_1 -1}  \frac{ \Gamma\Lb \omega+n\Rb}{\Gamma\Lb \omega\Rb\Gamma\Lb \omega_1\Rb}\Lb t\, G_{\pom}\Lb z'\Rb\Rb^n
\eeq
  After summation over $n$ we have
  \begin{subequations}
            \bea 
S\Lb z'\Rb &=&C^2  \!\!\!\! \intl^{\epsilon + i \infty}_{\epsilon - i \infty}\!\!\! \frac{ d \omega}{2\,\pi\,i} \!\!\!\intl^{\epsilon + i \infty}_{\epsilon - i \infty}\!\!\!  \frac{d \omega_1}{2\,\pi\,i} e^{ \frac{ \bg^2 \kappa}{2}\Lb (\omega-1)^2 + (\omega_1 -1)^2\Rb}\frac{ 1}{\Gamma\Lb \omega_1\Rb}\intl^\infty_0 d t\,  e^{-t} \Lb 1\,+\,t  \,G_{\pom}\Lb z'\Rb\Rb^{- \omega }\,t^{\omega_1 -1}\label{SA61}\\
 &\xrightarrow{G_{\pom}\Lb z'\Rb\,\gg\,1}&C^2 \!\!\!\! 
 \intl^{\epsilon + i \infty}_{\epsilon - i \infty}\!\!\! \frac{ d \omega}{2\,\pi\,i} \!\!\!\intl^{\epsilon + i \infty}_{\epsilon - i \infty}\!\!\!  \frac{d \omega_1}{2\,\pi\,i} e^{ \frac{ \bg^2 \kappa}{2}\Lb (\omega-1)^2 + (\omega_1 -1)^2\Rb}\intl^\infty_0 d t \, e^{-t} \exp\Lb -\omega\, \ln\Lb G_{\pom}\Lb z'\Rb\Rb\Rb\,\frac{t^{\omega_1 - \omega  -1}}{\Gamma\Lb \omega_1\Rb}\label{SA62}\\
 &=&C^2 \!\!\!\!
 \intl^{\epsilon + i \infty}_{\epsilon - i \infty}\!\!\! \frac{ d \omega}{2\,\pi\,i} \!\!\!\intl^{\epsilon + i \infty}_{\epsilon - i \infty}\!\!\!  \frac{d \omega_1}{2\,\pi\,i} e^{ \frac{ \bg^2 \kappa}{2}\Lb (\omega-1)^2 + (\omega_1 -1)^2\Rb}\frac{ \Gamma\Lb \omega_1 - \omega\Rb}{\Gamma\Lb \omega_1\Rb} \exp\Lb -\omega\, \ln\Lb G_{\pom}\Lb z'\Rb\Rb\Rb \label{SA63} 
\eea
  \end{subequations}
Closing contour of integration over  $\omega_1$ on the poles of $\Gamma\Lb \omega - \omega_1\Rb$  one can see that the main contribution gives the pole at $\omega = \omega_1$ while all other poles lead to the amplitude that decreases as $\exp\Lb -\bg\,n \,z'\Rb$ for the pole $\omega_1= \omega - n$. Taking integral over $\omega$ we obtain:
  \beq \label{SA7}
  S\Lb z'\Rb =  C'^2  \exp\Lb - \frac{ z'^2}{4 \kappa}\Rb
  \eeq
 Hence,  we can conclude that the sum of the large Pomeron loops leads to the  S-matrix in perfect agreement with the estimates of 'rare' fluctuation, given in the papers of Iancu and Mueller in  Ref.\cite{MPSI}.  $C'$ is a smooth function of $z$ which we cannot guarantee. Note that in our case it contains the factor $1/(1 + e^{\bg z})$ which could provide the matching at $z \to 0$ with the solution of Ref.\cite{MUT}.
  
\section{Multiplicity   distribution of produced gluons }

   \eq{SA5} is suited for using the AGK\cite{AGK} cutting rules for finding the multiplicity distributions of produced gluons. We recall that an application of AGK cutting rules is based on two main equations:
  
    \begin{subequations}
   \bea
2\, { \rm Im} \,G_{\pom}\Lb z \Rb\,\,&=&\,\,\sigma^{\mbox{\tiny BFKL}}_{in}(z); \label{AGK1}\\
 \sigma_n\Lb z \Rb\,\,&=&\,\,\sum_j\underbrace{ \sigma_j^{AGK}\Lb z \Rb}_{ \propto\,\Lb{\rm Im} G_{\pom} \Rb^j}\underbrace{ \frac{\Lb j \, \bg \,z\Rb^n}{n!} e^{ -j\, \bg\,z}}_{\mbox{Poisson distribition}} \label{AGK2}
 \eea
 \end{subequations}
   \eq{AGK1} is the $s$-channel unitarity constraint for the BFKL Pomeron, which clarifies the structure of the BFKL Pomeron. $\sigma^{\mbox{\tiny BFKL}}_{in}(z)$  is the inelastic cross sections of  produced gluons with mean multiplicity $\bar{n} = \bg \,z$  , where $\bg = 1 - \gamma_{cr}$ (see \eq{GACR}). We also know that produced gluons have the Poisson distribution  with this mean multiplicity\cite{LEDIDI}) . \eq{AGK2} gives the prescription how to calculate the inelastic cross section with given multiplicity using the AGK cutting rules. These rules, which we  discuss below allows us to find the 
 imaginary part of   the scattering amplitude, that determines the cross sections, through the powers of ${ \rm Im} \,G^{\mbox{\tiny BFKL}}\Lb z\Rb$\footnote{In the widespread slang this contribution is called by   cut Pomeron.}.  
 
 The AGK cutting rules \cite{AGK} allows us to calculate the contributions of $n$-cut Pomerons if we know $F_k$: the contribution of the exchange of $k$-Pomerons to  the cross section. They take the form:
   \begin{subequations} 
    \bea 
n\,\geq\,1:\sigma^k_n\Lb Y, \xi_{r,r'}\Rb&=& (-1)^{n -k}\frac{k!}{(n - k)!\,n!}\,2^{k}\, F_k(Y,\xi_{r,r'})\label{AGKK}\\
n\,=\,0:\sigma^k_0\Lb Y\Rb&=&\Lb -1\Rb^k \Bigg(2^k\,\,-\,\,2\Bigg) F_k(Y,\xi_{r,r'});\label{AGK0}\\
\sigma_{tot}&=&\,\,2 \sum_{k=1}^\infty (-1)^{k+1} \,F_k(Y,\xi_{r,r'});\label{XS}\,
\eea
 \end{subequations} 
$\sigma_{tot} = 2\, {\rm Im} A\Lb z\Rb$ where $A$ is the scattering amplitude. $\sigma_0$ is the cross section of diffractive  production of small numbers of gluons which is much smaller than  $ \bar{n} = \bg \,z$. 

 Using \eq{AGKK} and \eq{XS} we obtain from \eq{SA5}:
 
    \bea \label{AGK3}
&&\sigma^{\mbox{\tiny AGK}}_n\Lb z'\Rb \\
&& =C^2
\sum^\infty_{k=n} \frac{\Lb - 1\Rb^{k -n }}{k!}\Lb \frac{k!}{(k - n)! \,n!}\Rb\!\!\!  \intl^{\epsilon + i \infty}_{\epsilon - i \infty}\!\!\! \frac{ d \omega}{2\,\pi\,i} \!\!\!\intl^{\epsilon + i \infty}_{\epsilon - i \infty}\!\!\!  \frac{d \omega_1}{2\,\pi\,i} e^{ \frac{ \bg^2 \kappa}{2}\Lb (\omega-1)^2 + (\omega_1 -1)^2\Rb}\intl^\infty_0 d t e^{-t}  \frac{t^{\omega_1 -1} }{\Gamma\Lb \omega\Rb}\,\frac{\Gamma\Lb \omega + k\Rb}{\Gamma\Lb \omega_1\Rb}\Lb 2\,t\, G_{\pom}\Lb z'\Rb\Rb^k\nn\\
&& = \frac{C^2}{n!}\!\!\!  \intl^{\epsilon + i \infty}_{\epsilon - i \infty}\!\!\! \frac{ d \omega}{2\,\pi\,i} \!\!\!\intl^{\epsilon + i \infty}_{\epsilon - i \infty}\!\!\!  \frac{d \omega_1}{2\,\pi\,i} e^{ \frac{ \bg^2 \kappa}{2}
\Lb (\omega-1)^2 + (\omega_1 -1)^2\Rb}\intl^\infty_0  d t \, e^{-t}   \frac{t^{\omega_1 -1}}{\Gamma\Lb \omega\Rb\,\Gamma\Lb \omega_1\Rb} 
\Gamma\Lb n + \omega\Rb
\frac{\Lb 2\,t\, G_{\pom}\Lb z'\Rb\Rb^n }{\Lb  1+ 2\,t\, G_{\pom}\Lb z'\Rb\Rb^{ \omega + n}}
\eea

Integrating over $t$ we have
 \beq \label{AGK4}
\sigma^{\mbox{\tiny AGK}}_n\Lb z'\Rb  =\frac{C^2}{n!}\!\!\!  \intl^{\epsilon + i \infty}_{\epsilon - i \infty}\!\!\! \frac{ d \omega}{2\,\pi\,i} \!\!\!\intl^{\epsilon + i \infty}_{\epsilon - i \infty}\!\!\!  \frac{d \omega_1}{2\,\pi\,i} e^{ \frac{ \bg^2 \kappa}{2}\Lb (\omega-1)^2 + (\omega_1 -1)^2\Rb}\frac{\Gamma\Lb \omega + n\Rb\,\Gamma\Lb \omega_1 + n\Rb}{\Gamma\Lb \omega \Rb\,\Gamma\Lb \omega_1 \Rb}\,\frac{1}{N^{\omega_1}\Lb z' \Rb}\,U\Lb 
n+ \omega, \omega_1 - \omega +1,\frac{1}{N\Lb z'\Rb}\Rb
\eeq

where $N\Lb z'\Rb \,\,=\,\, 2\,\, G_{\pom}\Lb z'\Rb$  and  $U\Lb a,b,z\Rb$ is Tricomi confluent hypergeometric  function (see Ref.\cite{AS}, formula 13.1.3 ).  For large $N\Lb z'\Rb$ 
\beq \label{AGK5}
\Gamma\Lb \omega + n\Rb\,U\Lb 
n+ \omega, \omega_1 - \omega +1,\frac{1}{N\Lb z'\Rb}\Rb\xrightarrow{N\Lb z'\Rb \gg1} 2
\Lb \omega+n\Rb^{ -\Delta}
 \Lb \frac{1}{N\Lb z'\Rb}\Rb^{\Delta} \,K_{2\,\Delta}\Lb 2 \sqrt{\frac{n+ \omega}{N\Lb z'\Rb}}\Rb
\eeq
where $\Delta =\h\Lb \omega - \omega_1\Rb$.
For large $n$  \eq{AGK5} gives:
\beq \label{AGK6}
\Gamma\Lb \omega + n\Rb\,U\Lb 
n+ \omega, \omega_1 - \omega +1,\frac{1}{N\Lb z'\Rb}\Rb\xrightarrow{n > N\Lb z'\Rb } 2 \Lb \omega+n\Rb^{- \Delta-1/4}\Lb \frac{1}{N\Lb z'\Rb}\Rb^{-\Delta-1/4} \,\exp\Lb - \Lb 2 \sqrt{\frac{n+ \omega}{N\Lb z'\Rb}}\Rb\Rb
\eeq

 The expression for $\sigma_n^{\mbox{\tiny AGK}} $ takes the form:
 \beq \label{AGK6} 
 \sigma^{\mbox{\tiny AGK}}_n\Lb z'\Rb  =\frac{C^2}{n!}\!\!\!  \intl^{\epsilon + i \infty}_{\epsilon - i \infty}\!\!\! \frac{ d \omega}{2\,\pi\,i} \!\!\!\intl^{\epsilon + i \infty}_{\epsilon - i \infty}\!\!\!  \frac{d \omega_1}{2\,\pi\,i} e^{ \frac{ \bg^2 \kappa}{2}\Lb (\omega-1)^2 + (\omega_1 -1)^2\Rb}\frac{\Gamma\Lb \omega_1 + n\Rb}{\Gamma\Lb \omega \Rb\,\Gamma\Lb \omega_1 \Rb}\,\frac{\Lb \omega + n\Rb^{- \Delta-1/4}}{N^{\Sigma -1/4}\Lb z' \Rb}2  \,\exp\Lb - \Lb 2 \sqrt{\frac{n+ \omega}{N\Lb z'\Rb}}\Rb\Rb \eeq 
 
  Taking the integral using the method of steepest descent, we can see that $\omega_1  \sim \ln N\Lb z'\Rb$ is large. Hence \eq{AGK6}  for $ n \,<\,\omega_1$ can be rewritten as
   \beq \label{AGK7} 
\hspace{-0.3cm} \sigma^{\mbox{\tiny AGK}}_n\Lb z'\Rb  =\frac{C^2}{n!}\!\!\!  \intl^{\epsilon + i \infty}_{\epsilon - i \infty}\!\!\! \frac{ d \Sigma}{2\,\pi\,i} \!\!\!\intl^{\epsilon + i \infty}_{\epsilon - i \infty}\!\!\!  \frac{d \Delta}{2\,\pi\,i} e^{  \bg^2 \kappa\Lb (\Sigma-1)^2 \,+\,  \Delta^2\Rb}  \frac{\Lb \Sigma + \Delta\Rb^{n}}{\Gamma\Lb \Sigma + \Delta \Rb\,}\,\frac{2\Lb n+ \Sigma + \Delta\Rb^{-\Delta-1/4}}{N^{\Sigma-1/4}\Lb z' \Rb}  \,\exp\Lb - \Lb 2 \sqrt{\frac{n+ \Sigma + \Delta}{N\Lb z'\Rb}}\Rb\Rb \eeq  
 where $\Sigma= \h\Lb \omega + \omega_1\Rb$.
 
  The equations for saddle points have the following forms, assuming that $\Delta^{SP} \ll \Sigma^{SP}$:
     \begin{subequations}   
     \bea
  &&  2\, \bg^2 \kappa\,\Lb \Sigma^{SP} -1\Rb -  \ln N\Lb z'\Rb - \ln \Sigma^{SP}=\,0; \label{AGKSPOM}\\
  &&
  2\, \bg^2 \kappa\, \Delta^{SP}   -  \ln \Lb n + \Sigma^{SP}\Rb\,=\,0;\label{AGKSPDE}
 \eea
 \end{subequations}   
 The solutions are
   \begin{subequations}   
     \bea
  &&  \Sigma^{SP} \,-\,1 \,=\,\frac{1}{2\,\bg^2\,\kappa}\Bigg\{ \ln N\Lb z'\Rb + \ln \Lb\frac{1}{2\,\bg^2\,\kappa} \ln N\Lb z'\Rb \Rb\ \Bigg\}; \label{AGKSPOMS}\\
  &&
 \Delta^{SP}   =  \frac{1}{  2\, \bg^2 \kappa}   \ln \Lb n + \Sigma^{SP}\Rb;\label{AGKSPDES}
 \eea     
  \end{subequations}  
  
  From these equations one can see that for $n > N\Lb z'\Rb$ $\Delta^{SP}$ starts to exceed $\Sigma^{SP}$ . Therefore we will consider this case separately below. Using \eq{AGKSPOMS} and \eq{AGKSPDES}
  we integrate over $\Sigma$ and $\Delta$ and find $\sigma_n$ in the first region:    
  
  \begin{boldmath}
I.  $n \,<\,\frac{1}{2\,\bg^2\,\kappa}\ln N\Lb z'\Rb  \equiv \frac{1}{2\,\bg^2\,\kappa}  \zeta$
  \end{boldmath}

      In this region $\sigma_n$ are equal:
         \beq \label{AGK8} 
 \sigma^{\mbox{\tiny AGK}}_n\Lb z'\Rb  =\frac{C^2\,e^{-\frac{3}{4}\zeta}}{n!}\,  e^{  - \frac{1}{4\,\bg^2\,\kappa}\zeta^2 } \frac{\Lb \frac{1}{2\,\bg^2\,\kappa}\zeta\Rb^{n}}{\Gamma\Lb \frac{1}{2\,\bg^2\,\kappa}\zeta \Rb\,}\, \,\exp\Lb -  2 \sqrt{\frac{\frac{1}{2\,\bg^2\,\kappa}\zeta}{N\Lb z'\Rb}}\Rb \eeq  
     Therefore, in this region the produced gluons have  the Poisson distribution with the mean multiplicity $\bar{n}^{I}\,=\, \frac{1}{2\,\bg^2\,\kappa}\zeta$.     
  
  We have to return to \eq{AGK6} to find $\sigma_n$ in the second region:
  
    \begin{boldmath}
II.  $N\Lb z' \Rb\,\,>\,n \,>\,\,\frac{1}{2\,\bg^2\,\kappa}\ln N\Lb z'\Rb  \equiv \frac{1}{2\,\bg^2\,\kappa}  \zeta$
  \end{boldmath}
   
   For $n > \omega_1$ \eq{AGK6} takes the form:
   
       \bea \label{AGK9} 
&& \sigma^{\mbox{\tiny AGK}}_n\Lb z'\Rb  =\\
&&C^2\!\!\!  \intl^{\epsilon + i \infty}_{\epsilon - i \infty}\!\!\! \frac{ d \Sigma}{2\,\pi\,i} \!\!\!\intl^{\epsilon + i \infty}_{\epsilon - i \infty}\!\!\!  \frac{d \Delta}{2\,\pi\,i} e^{  \bg^2 \kappa\Lb (\Sigma -1)^2 \,+\,  \Delta^2\Rb}  \frac{n^{ \Sigma + \Delta -1}}{\Gamma\Lb \Sigma + \Delta \Rb\,\Gamma\Lb \Sigma - \Delta \Rb\,}\,\frac{2}{N^{\Sigma -1/4}\Lb z' \Rb} \Lb n+ \Sigma + \Delta\Rb^{-\Delta-1/4} \,\exp\Lb - \Lb 2 \sqrt{\frac{n+ \Sigma + \Delta}{N\Lb z'\Rb}}\Rb\Rb\nn \eea   
     
     For \eq{AGK9} 
 the  saddle points for  $\Delta^{SP} \ll \Sigma^{SP}$ satisfy the following equations:
     \begin{subequations}   
     \bea
  &&  2\, \bg^2 \kappa\, \Lb \Sigma^{SP}-1\Rb  -  \ln N\Lb z'\Rb - 2\ln \Sigma^{SP} \,+\,\ln n=\,0; \label{AGKSPOM1}\\
  &&
  2\, \bg^2 \kappa\, \Delta^{SP}   +   \ln \Lb n + \Sigma^{SP}\Rb\,=\,0;\label{AGKSPDE1}
 \eea
 \end{subequations}  
with solutions:
   \begin{subequations}   
     \bea
  &&  \Sigma^{SP} -1 \,=\,\frac{1}{2\,\bg^2\,\kappa}\Bigg\{ \ln N\Lb z'\Rb +  2\,\ln \Lb\frac{1}{2\,\bg^2\,\kappa} \ln N\Lb z'\Rb \Rb\, - \,\ln n \Bigg\}; \label{AGKSPOMS1}\\
  &&
 \Delta^{SP}   = - \frac{1}{  2\, \bg^2 \kappa}   \ln \Lb n + \Sigma^{SP}\Rb;\label{AGKSPDES1}
 \eea     
  \end{subequations}

   Taking integrals we obtain:
     \beq \label{AGK10} 
 \sigma^{\mbox{\tiny AGK}}_n\Lb z'\Rb  =
C^2\,e^{-\zeta} \,\frac{1}{n}\,\Lb \frac{N\Lb z \Rb}{n}\Rb^{1/4} \exp\Bigg\{ - \frac{1}{4\,\bg^2\,\kappa}\Lb \zeta + 2 \ln \Lb \frac{\zeta}{2 \,\bg^2\,\kappa}\Rb - \ln n \Rb^2 \,-\, \frac{1}{4\,\bg^2\,\kappa}\ln^2 n \Bigg\} \, \,\exp\Bigg( -  2 \sqrt{\frac{n}{N\Lb z'\Rb}} \Bigg) \eeq  
       
 In the region III:
  
    \begin{boldmath}
III.  $n\,>\, N\Lb z' \Rb$
  \end{boldmath}
   
    $\Delta \,\gg \,\Sigma$    and using \eq{AGK9} we obtain the equations for the saddle points:
     For \eq{AGK9} 
 the  saddle points for  $\Delta^{SP} \ll \Sigma^{SP}$ satisfy the following equations:
     \begin{subequations}   
     \bea
  &&  2\, \bg^2 \kappa\,\Lb \Sigma^{SP} -1\Rb -  \ln N\Lb z'\Rb =\,0; \label{AGKSPOM1}\\
  &&
  2\, \bg^2 \kappa\, \Delta^{SP}   +  2 \ln \Lb\frac{ n}{\Delta^{SP}}\Rb\,=\,0;\label{AGKSPDE1}
 \eea
 \end{subequations}  
They have solutions 
   \begin{subequations}   
     \bea
  &&  \Sigma^{SP} -1 \,=\,\frac{1}{2\,\bg^2\,\kappa}\ln N\Lb z'\Rb = \frac{1}{2\,\bg^2\,\kappa}\zeta; \label{AGKSPOMS2}\\
  &&
 \Delta^{SP}   = - \frac{1}{ \ \bg^2 \kappa}   \ln \Lb\frac{ n}{ \frac{1}{ \ \bg^2 \kappa}\ln n} \Rb;\label{AGKSPDES2}
 \eea     
  \end{subequations}
  Taking the integrals we have:
  
  \beq \label{AGK11} 
 \sigma^{\mbox{\tiny AGK}}_n\Lb z'\Rb  =C^2\,e^{-\zeta} \,\frac{1}{n}\,\Lb \frac{N\Lb z \Rb}{n}\Rb^{1/4} \exp\Bigg\{ - \frac{1}{4\,\bg^2\,\kappa}\Lb  \zeta^2 \, +  \ln^2 \Lb\frac{ n}{ \frac{1}{\bg^2 \kappa}\ln n} \Rb\Rb\Bigg\} \, \,\exp\Bigg( -  2 \sqrt{\frac{n}{N\Lb z'\Rb}} \Bigg) \eeq

    Now we are going to calculate the total inelastic cross section:
 \beq \label{AGL12}
   \sigma_{in}\,\,=\,\,\sum^{\infty}_{n=1} \sigma_n\,\,=\,\,\sum^{\infty}_{n=1} \sigma^{AGK}_n    
    \eeq
  We return to \eq{AGK3} and sum over $n$ which reduces this equation to the form:
    \beq \label{AGL13} 
    \sigma_{in} = C^2 \intl^{\epsilon + i \infty}_{\epsilon - i \infty}\!\!\! \frac{ d \omega}{2\,\pi\,i} \!\!\!\intl^{\epsilon + i \infty}_{\epsilon - i \infty}\!\!\!  \frac{d \omega_1}{2\,\pi\,i} e^{ \frac{ \bg^2 \kappa}{2}
    \Lb( \omega-1)^2 + (\omega_1-1)^2\Rb}\intl^\infty_0  d t \, e^{-t}   \frac{t^{\omega_1 -1}}{\Gamma\Lb \omega_1\Rb} \Bigg\{ 1\,\,-\,\,\frac{1}{\Lb 1\,+\,t\,N\Lb z'\Rb\Rb^\omega}\Bigg\}
 \eeq     
    Integrating over $t$ leads to the following expression:
    \beq \label{AGL14} 
    \sigma_{in} = C^2 \intl^{\epsilon + i \infty}_{\epsilon - i \infty}\!\!\! \frac{ d \omega}{2\,\pi\,i} \!\!\!\intl^{\epsilon + i \infty}_{\epsilon - i \infty}\!\!\!  \frac{d \omega_1}{2\,\pi\,i} e^{ \frac{ \bg^2 \kappa}{2}
 \Lb( \omega-1)^2 + (\omega_1-1)^2\Rb } \Bigg\{ 1\,\,-\,\,\,\Lb\frac{1}{N\Lb z'\Rb }\Rb^{\omega_1}U\Lb \omega_1,  \omega_1  - \omega, \frac{1}{N\Lb z'\Rb}\Rb\Bigg\}
 \eeq   
 Using \eq{AGK6} we obtain:
\beq \label{AGK15} 
 \sigma_{in}\Lb z'\Rb  =C^2 \intl^{\epsilon + i \infty}_{\epsilon - i \infty}\!\!\! \frac{ d \Sigma}{2\,\pi\,i} \!\!\!\intl^{\epsilon + i \infty}_{\epsilon - i \infty}\!\!\!  \frac{d \Delta}{2\,\pi\,i} e^{  \bg^2 \kappa\Lb (\Sigma-1)^2 \,+\,  \Delta^2\Rb}\Bigg\{1\,-\,  \frac{1}{\Gamma\Lb \Sigma + \Delta \Rb}\,\frac{2}{N^{\Sigma}\Lb z' \Rb} \Lb\Sigma + \Delta\Rb^{\Delta} \,\exp\Lb - 2 \sqrt{\frac{\Sigma + \Delta}{N\Lb z'\Rb}}\Rb \Bigg\}\eeq   
    
    Taking integrals over $\Sigma$ and $\Delta$ using the method of steepest descent we have:
    \beq \label{AGK16} 
 \sigma_{in}\Lb z'\Rb  =C^2\Bigg\{1\,-\,  e^{-\zeta} e^{ - \frac{1}{4\,\bg^2\,\kappa}\Lb \zeta + \ln\Lb \frac{\zeta}{2\,\bg^2\,\kappa}\Rb\Rb^2  - \frac{1}{4\,\bg^2\,\kappa} \ln^2\Lb \frac{\zeta}{2\,\bg^2\,\kappa}\Rb} \,\exp\Lb - 2 \sqrt{\frac{\frac{1}{4\,\bg^2\,\kappa} \zeta}{N\Lb z'\Rb}}\Rb \Bigg\}\eeq  
   For the cross section of the inclusive production of the produced gluons we have
     \bea \label{AGK17} 
 \sigma_{incl}\Lb z'\Rb \,\,&=&\,\,\sum^{\infty}_{n=1}n\, \sigma^{AGK}_n = C^2 \intl^{\epsilon + i \infty}_{\epsilon - i \infty}\!\!\! \frac{ d \Sigma}{2\,\pi\,i} \!\!\!\intl^{\epsilon + i \infty}_{\epsilon - i \infty}\!\!\!  \frac{d \Delta}{2\,\pi\,i} e^{  \bg^2 \kappa\Lb( \Sigma-1)^2 \,+\,  \Delta^2\Rb} \omega \intl^\infty_0 d t \,e^{-t}N\Lb z'\Rb \frac{t^{\omega_1}}{\Gamma\Lb \omega_1\Rb}\,\,\nn\\ 
& =& C^2 \intl^{\epsilon + i \infty}_{\epsilon - i \infty}\!\!\! \frac{\omega\, d \omega}{2\,\pi\,i} \!\!\!\intl^{\epsilon + i \infty}_{\epsilon - i \infty}\!\!\!  \frac{\omega_1\,d \omega_1}{2\,\pi\,i} e^{  \bg^2 \kappa\Lb( \omega-1)^2 \,+\,  (\omega_1-1)^2\Rb}  = 
C^2 \,N\Lb z'\Rb
 \eea
    One can see that the integrals over $\omega$ and $\omega_1$ would lead to zero if we do not put $\omega -1$ and $\omega_1 - 1$ in the exponent. The defect of the expression for the dipole densities in Eq.12a of Ref.\cite{LEDIDI} is that  in the series of this equation the contribution of one Pomeron exchange vanishes due to $\omega$ integration. Actually \eq{AGK17} follows directly from the AGK cutting rules at any value of $z$ with ${\rm C^2}=1$. Note, that the multiplicity distribution $\mathcal{P}_n\,=\,\sigma_n/\sigma_{in}$ does not depend on the value of the smooth function ${\rm C}$.
    
    We introduce the ration $\mathcal{P}^{gluon}_n= \frac{ \sigma^{\mbox{\tiny AGK}}_n\Lb z'\Rb}{ \sigma^{\mbox{\tiny AGK}}_{\rm  incl}\Lb z'\Rb}  $ to get rid of the unknown smooth function $C^2$ which is equal to 
        \beq \label{AGK18} 
\mathcal{P}^{gluon}_n= \frac{ \sigma^{\mbox{\tiny AGK}}_n\Lb z'\Rb}{ \sigma^{\mbox{\tiny AGK}}_{\rm  incl}\Lb z'\Rb}  =\frac{1}{n} \frac{1}{N\Lb z \Rb}\Lb \frac{n}{N\Lb z \Rb}\Rb^{-1/4}\!\!\!\!\exp\Bigg\{ - \frac{1}{4\,\bg^2\,\kappa}\Lb  \zeta^2 \, +  \ln^2 \Lb\frac{ n}{ \frac{1}{\bg^2 \kappa}\ln n} \Rb\Rb\Bigg\} \, \,\exp\Bigg( -  2 \sqrt{\frac{n}{N\Lb z'\Rb}} \Bigg) \eeq     
     
\section{Inclusive production of $J/\Psi$ in the BFKL cascade }

\begin{figure}[h]
\centering \leavevmode \includegraphics[width=12cm]{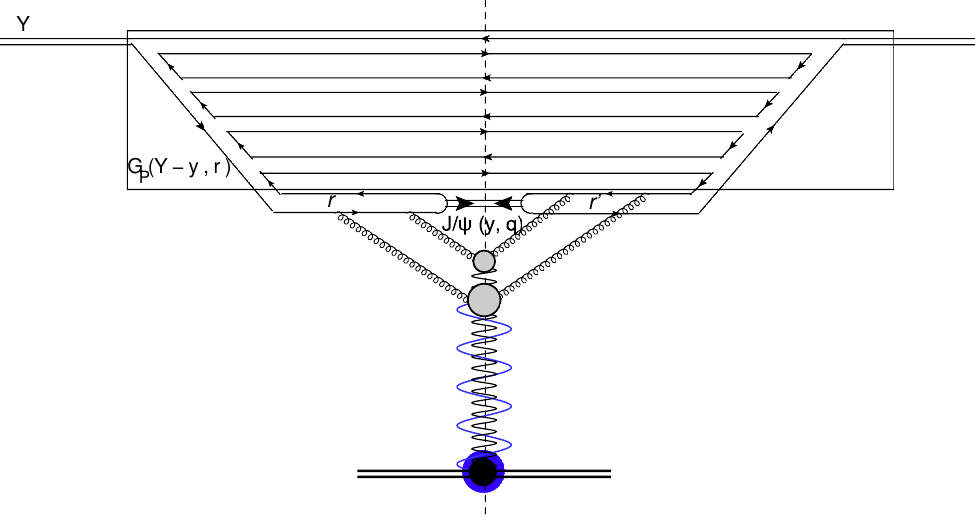}
\caption{The cross-section corresponding to the first diagram of the \fig{3p}-a
in the BFKL Pomeron calculus.The vertical wavy lines of different
colors and shape passing through unitarity cut are BFKL Pomerons (described
by the Green functions $G_{\pom}\Lb Y_{J/\Psi}, r\Rb $ or $G_{\pom}\Lb Y_{J/\Psi}, r' \Rb$  i n \eq{FD11})), helical lines correspond to the gluons. }
\label{fidi} 
\end{figure}

     First,  we rewrite in our notations the cross section of $J/\Psi$ production due to three gluon fusion that has been written in Ref.\cite{LESI}. This cross section is shown in \fig{3p}-a and in \fig{fidi}, where are  shown our main notations. For simplicity we consider the inclusive cross section integrated over the transverse momentum $q_T$ of produced $J/\Psi$. This cross section is equal to (see Eq.9 of Ref.\cite{LESI}):
     \bea\label{FD11} 
     && \frac{d\sigma_{J/\Psi}\left(y\right)}{dy}\,\,=\,\,\,2\,\int_{0}^{1}dz\int_{0}^{1}dz'\int\frac{d^{2}r}{4\pi}\,\frac{d^{2}r'}{4\pi} \Lb 2\,N_\pom\Lb Y - y, r\Rb \Rb\,\\
 &&\Psi_{J/\psi}\left(r , z\right)\,\Psi_{J/\psi}\left(r',z'\right)
 \Bigg(\,2\,N\left(y;\,\frac{\vec{r}+\vec{r}'}{2}\right)\,-\,2\,N\left(y;\,\frac{\vec{r}-\vec{r}'}{2}\right)\Bigg)^{2}\nonumber 
\eea  
  
   The wave function of $J/\Psi$ can be found in the appendix to Ref.\cite{LESI}    as well as other details of the derivation. One can recognize that $2 \,N$ in this equation corresponds to the unitarity constraints of \eq{AGL12}. We can simplify \eq{FD11} by integrating over angle between  $\vec{r}$ and $\vec{r}'$ using that $r$=$r'$ with good accuracy since both are equal to $r_d$: the typical size of  $c\,\bar{c}$ -dipole of the $J/\Psi$ meson.  Using \eq{DD1} for $N$ we obtain:
    
     \begin{subequations}     
    \bea 
     && \frac{d\sigma_{J/\Psi}\left(y\right)}{dy}\,\,=\,\,\,2\,\int_{0}^{1}dz\int_{0}^{1}dz'\int\frac{d^{2}r}{4\pi}\,\frac{d^{2}r'}{4\pi} \Lb 2\,N_\pom\Lb Y - y, r\Rb \Rb\,\label{FD12}\\
 &&\Psi_{J/\psi}\left(r , z\right)\,\Psi_{J/\psi}\left(r',z'\right)\,4 Q^{2 \bg}\Lb Y - y\Rb \Lb 4 \sqrt{\pi} \Lb 2^{4 \bg}\frac{ \Gamma\Lb \h + 4 \bg\Rb}{\Gamma\Lb 1 + 4 \bg\Rb} \,-\,\frac{ \Gamma\Lb \h + 2 \bg\Rb}{\Gamma\Lb 1 + 2 \bg\Rb}\Rb\Rb \Lb r^2\,r'^2\Rb^{\bg}\nonumber \\
 && {\rm Const}\,\, G^{\rm cut}_\pom\Lb z_d\Rb \Lb G^{\rm cut}_\pom\Lb z'_d\Rb\Rb^2\label{FD13}  \eea   
      \end{subequations}

     $G^{\rm cut}_\pom\Lb z_d\Rb $ and $G^{\rm cut}_\pom\Lb z'_d\Rb$ denote the Green's function of the cut BFKL Pomeron.
In \eq{FD13} we do not specify the value of the {\rm Const} since we cannot guarantee this value in our approach. $z_d$ and $z'_d$ are equal to
     \beq \label{zd}
z_d =  \,\,\,\,\bas \frac{\chi\Lb \bg\Rb}{\bg} \,\Lb Y - y\Rb \,\,+\,\,\xi_{r,r_d} ;   ~~~~~     
z'_d =  \,\,\,\,\bas \frac{\chi\Lb \bg\Rb}{\bg} \,y \,\,+\,\,\xi_{r_d,r'} ; ~~~~~     
z =  \,\,\,\,\bas \frac{\chi\Lb \bg\Rb}{\bg} \,Y \,\,+\,\,\xi_{r,r'} ;
\eeq

  Using our approach of Re.\cite{LEDIDI} we can rewrite \eq{FD13} as the first diagram of the BFKL Parton cascade for $\sigma_n$  with produced $J/\Psi$:
  \bea \label{FD14}
&&{\rm Const}\,\, 2 \,\rho_1\Lb z_d\Rb \, 4 \rho_1\Lb z'_{d,i}\Rb \,\rho_1\Lb z'
_{d,j} \Rb\,\,\equiv \,\,{\rm Const}\,\, 2 \,\rho_1\Lb z_i\Rb \, 4  \,\rho_1\Lb z' 
_{d,j} \Rb;\\
 &&\mbox{with}  ~~~z_{d,i} =  \,\,\,\,\bas \frac{\chi\Lb \bg\Rb}{\bg} \,y \,\,+\,\,\xi_{r_d,r_i};\,\,
 z_{d,j} =  \,\,\,\,\bas \frac{\chi\Lb \bg\Rb}{\bg} \,y \,\,+\,\,\xi_{r_d, r_j} ;    \,\,
 z_{i} =  \,\,\,\,\bas \frac{\chi\Lb \bg\Rb}{\bg} \,Y \,\,+\,\,\xi_{r, r_i} ;  \nn  
 \eea
Note, that $\rho_1$ in this equation is given by \eq{RHO1}.

\begin{figure}[ht]
    \centering
  \leavevmode
      \includegraphics[width=0.7\textwidth]{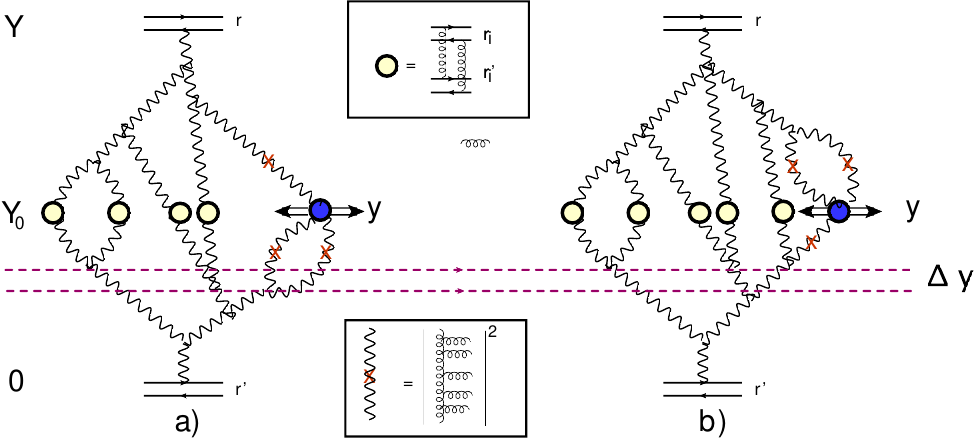} 
     \caption{\fig{mpsipsi}: Summing  large Pomeron loops for $J/\Psi$ production in  dipole-dipole scattering.           
The wavy lines denote the  BFKL Pomeron exchanges.  The circles denote the amplitude $\gamma$ in the Born approximation of perturbative QCD. The blue blob denoted the $J/\Psi$ production due to three gluons fusion. The wavy lines with the cross indicate the cut Pomerons. The dashed lines denote the rapidity window, where the multiplicity of produced gluons is measured.  The picture reflects the Mueller diagram technique for inclusive production (see Ref.\cite{MUDIA}).}
\label{mpsipsi}     
   \end{figure}

 The general structure of the $J/\Psi$ production in our approach is shown in \fig{mpsipsi}. Choosing $Y_0 =y$ in \eq{MPSI} we see from \fig{mpsipsi} that we can   calculate the inclusive production of $J/\Psi$ using \eq{MPSI} for the scattering amplitude but introducing a new vertex (blue circle in \fig{mpsipsi}) for the $J/\Psi$ production and taking into account that in \eq{MPSI} at least one Pomeron is a cut Pomeron.
      Bearing this in mind we can rewrite \eq{DD1} for the projectile densities in the following form for \fig{mpsipsi}-a:
      
\beq \label{DA1}
 \rho^P_n\Lb Y -y;   r,b, \{r_i,b_i\}\Rb   \,\,=\,\, 
 \frac{1}{n!} \intl^{\epsilon + i \infty}_{\epsilon - i \infty}\!\!\!  \frac{d \omega_1}{2\,\pi\,i} 
e^{ \frac{ \bg^2 \kappa}{2} (\omega_1-1)^2 }\frac{ \Gamma\Lb \omega_1+n\Rb}{\Gamma\Lb \omega_1\Rb} \,\prod_{i=1}^{n} G_{\pom}\Lb z_{d,j}\Rb \eeq  

For the target densities we have:
\beq \label{DA2}
 \rho^T_n\Lb y;   r,b, \{r_i,b_i\}\Rb   \,\,=\,\, 
 \frac{1}{(n+1)!} \intl^{\epsilon + i \infty}_{\epsilon - i \infty}\!\!\!  \frac{d \omega_2}{2\,\pi\,i} 
e^{ \frac{ \bg^2 \kappa}{2} (\omega_2-1)^2 }\frac{ \Gamma\Lb \omega_2+n+1\Rb}{\Gamma\Lb \omega_2\Rb} \,\prod_{i=1}^{n+1} G_{\pom}\Lb z_{d,j}\Rb \eeq

\section{Multiplicity distribution for  $J/\Psi$ production  in dipole-dipole scattering }
         Using \eq{DD5} and  \eq{DA2} for the dipole densities we can calculate the scattering amplitude for the $J/\Psi$ production from \eq{MPSI},viz.:
    \bea \label{MDJ1}
&&   S_{J/\Psi} \Lb z',z'_d\Rb     =C^2\sum^\infty_{n=0} \frac{\Lb - 1\Rb^n}{n!}\!\!\!\intl^{\epsilon + i \infty}_{\epsilon _ i \infty}\!\!\! \!\! \frac{d \omega_1}{2\,\pi\,i}\!\!\!\intl^{\epsilon + i \infty}_{\epsilon _ i \infty}\!\!\! \!\! \frac{d \omega_2}{2\,\pi\,i} e^{ \frac{ \bg^2 \kappa}{2}\Lb( ( \omega_1-1) + (\omega_2-1)^2\Rb}\frac{ \Gamma\Lb \omega+n\Rb}{\Gamma\Lb \omega\Rb}\\
&\times&{\Gamma\Lb\omega_2+ 1 +n\Rb}{\Gamma\Lb \omega_1\Rb}\frac{ t^{\omega_2 - 1 + n - k}_2}{\Gamma\Lb \omega_2\Rb} \prod^k_{i=1} \intl\!\!\! d^2 r_i d^2 r'_i d^2 b_i d^2 b'_i\prod^{n-1}_{i=1} 
G_{\pom}\Lb z_{d,i}\Rb\gamma^{BA}\Lb r_i,r'_i,\delta b\Rb\, G_{\pom}\Lb z_{d,j}\Rb G*2_{\pom}\Lb z'_{d,j}\Rb\nn\eea
We wish to emphasize that this is our main equation for inclusive production. In this equation we assume that 
$Y_0 =y$ is the central region of rapidity and we can use the dipole densities of \eq{DA1} and \eq{DA2}.

   Making integrations using \eq{TU2} we obtain:
           
          \beq \label{MDJ2}
  S_{J/\Psi} \Lb z',z'_d\Rb     =C^2\sum^\infty_{n=0} \frac{\Lb - 1\Rb^n}{n!} \!\!\!\intl^{\epsilon + i \infty}_{\epsilon _ i \infty}\!\!\! \!\! \frac{d \omega_1}{2\,\pi\,i}\!\!\!\intl^{\epsilon + i \infty}_{\epsilon _ i \infty}\!\!\! \!\! \frac{d \omega_2}{2\,\pi\,i} e^{ \frac{ \bg^2 \kappa}{2}\Lb( \omega_1-1) + (\omega_2-1)^2\Rb}\,
\frac{\Gamma\Lb \omega_1 + n\Rb}{\Gamma\Lb \omega_1\Rb}\frac{ \Gamma\Lb \omega_2 + n + 1\Rb}{\Gamma\Lb \omega_2\Rb}\,G^n_{\pom}\Lb z\Rb\, G_{\pom}\Lb z'_{d}\Rb
\eeq     
We use the AGK cutting rules (see \eq{AGK3} and \eq{AGKK}-\eq{XS}) to find the cross section for $k$ cut Pomerons with the Green's functions  $G^k_{\pom}\Lb z'\Rb$ ($\sigma^{{\rm \tiny AGK}}_k\Lb z'\Rb$)  and for $n-k$ cut Pomerons the Green's functions  $G^{n-k}_{\pom}\Lb z'_s\Rb$ ($\sigma^{{\rm \tiny AGK}}_{n-k}\Lb z'\Rb$). For this goal we have to consider the scattering amplitude with $l $ Pomerons in $t$-channel (see \eq{MPSI}) and apply the AGK cutting rules to this amplitude. The expression for 
 $ \sigma^{\mbox{\tiny AGK}}_n\Lb z',z'_d\Rb$ \footnote{ We hope that   using the same notation as in section III $\sigma^{\mbox{\tiny AGK}}_n\Lb z',z'_d\Rb$ will be not confusing since we have different arguments.}  takes the form 
        \bea \label{MDJ3}
&& \sigma^{\mbox{\tiny AGK}}_n\Lb z',z'_d\Rb          =\\
&&C^2 \frac{1}{n!}\!\!\!\intl^{\epsilon + i \infty}_{\epsilon _ i \infty}\!\!\! \!\! \frac{d \omega_1}{2\,\pi\,i}\!\!\!\intl^{\epsilon + i \infty}_{\epsilon _ i \infty}\!\!\! \!\! \frac{d \omega_2}{2\,\pi\,i} e^{ \frac{ \bg^2 \kappa}{2}\Lb( \omega_1-1)^2 + (\omega_2-1)^2\Rb}\,n\,  \frac{\Gamma\Lb \omega_1 + l\Rb}{\Gamma\Lb \omega_1\Rb} \underbrace{\sum^\infty_{l=k} \frac{(-1)^{l-n}}{l!} \underbrace{\frac{l!}{(l - n)!\,n!}  \Lb 2 \,G_{\pom}\Lb z\Rb\Rb^l  2 \,G_{\pom}\Lb z_d\Rb}_{AGK \,cutting\,rules\,}}_{ \sigma^{\mbox{\tiny AGK}}_k\Lb z'\Rb(see\,\eq{MPSI})}\frac{\Gamma\Lb \omega_2 + l+1\Rb}{\Gamma\Lb \omega_1\Rb}\nn
\eea  

The extra factor of $n$ in \eq{MDJ3} reflects the possibility to produce $J/\Psi$ from each of $n$ BFKL Pomerons.

Making  summations   over $l$  and replacing $\Gamma\Lb \omega_1+l\Rb$ by \eq{DA21}  we have:
        \bea \label{MDJ4}
\hspace{-0.3cm} \sigma^{\mbox{\tiny AGK}}_n\Lb z',z'_d\Rb    &=&\frac{C^2}{(n-1)!}\!\!\!\intl^{\epsilon + i \infty}_{\epsilon _ i \infty}\!\!\! \!\! \frac{d \omega_1}{2\,\pi\,i}\!\!\!\intl^{\epsilon + i \infty}_{\epsilon _ i \infty}\!\!\! \!\! \frac{d \omega_2}{2\,\pi\,i} e^{ \frac{ \bg^2 \kappa}{2}\Lb( ( \omega_1-1) + (\omega_2-1)^2\Rb}\frac{\Gamma\Lb \omega_2 + n+1\Rb}{\Gamma\Lb \omega_1\Rb\,\Gamma\Lb \omega_2\Rb}
 \,\intl^\infty_0\!\!e^{- t} d t\,t^{\omega_1 - 1 }
\frac{ \Lb t\, N\Lb z\Rb\Rb^n N\Lb z'_d\Rb}{\Lb 1\,\,+\,\,t N\Lb z \Rb\Rb^{\omega_2 + n +1}}\\
&=&\frac{C^2 N\Lb z'_d\Rb}{(n-1)!}\!\!\!\intl^{\epsilon + i \infty}_{\epsilon _ i \infty}\!\!\! \!\! \frac{d \omega_1}{2\,\pi\,i}\!\!\!\intl^{\epsilon + i \infty}_{\epsilon _ i \infty}\!\!\! \!\! \frac{d \omega_2}{2\,\pi\,i} e^{ \frac{ \bg^2 \kappa}{2}\Lb( ( \omega_1-1) + (\omega_2-1)^2\Rb}\frac{\Gamma\Lb \omega_2 + n+1\Rb\Gamma\Lb \omega_1+n\Rb}{\Gamma\Lb \omega_1\Rb\,\Gamma\Lb \omega_2\Rb}\Lb \frac{1}{N\Lb z\Rb}\Rb^{\omega_1}  \!\!\!\!\! \!U\Lb \omega_1+n,\omega_1 - \omega_2, \frac{1}{N\Lb z \Rb}\Rb\nn
\eea 

where $N\Lb z\Rb = 2 G_\pom\Lb z\Rb$ and $N \Lb z'_d\Rb = 2 G_\pom\Lb z'_d\Rb$.

Using \eq{AGK5} and \eq{AGK6} we reduce \eq{MDJ4} to the form:

        \bea \label{MDJ5}
\sigma^{\mbox{\tiny AGK}}_n\Lb z',z'_d\Rb&=& \frac{C^2 N\Lb z'_d\Rb}{(n-1)!}\!\!\!\intl^{\epsilon + i \infty}_{\epsilon _ i \infty}\!\!\! \!\! \frac{d \Sigma}{2\,\pi\,i}\!\!\!\intl^{\epsilon + i \infty}_{\epsilon _ i \infty}\!\!\! \!\! \frac{d \Delta}{2\,\pi\,i} e^{ \bg^2 \kappa\Lb( ( \Sigma - 1)^2 + \Delta^2\Rb}\frac{\Gamma\Lb \Sigma - \Delta + n+1\Rb}{\Gamma\Lb \Sigma + \Delta\Rb\,\Gamma\Lb \Sigma - \Delta\Rb}\nn\\
&\times&\Lb \frac{1}{N\Lb z\Rb}\Rb^{\Sigma - 1/4}\Lb \Sigma - \Delta+n -1\Rb^{\Delta -1/4}\exp\Lb -2 \sqrt{\frac{\Sigma + \Delta +n -1}{N\Lb z\Rb}}\Rb
\eea
In the region 

 \begin{boldmath}
I.  $n \,<\,\frac{1}{2\,\bg^2\,\kappa}\ln N\Lb z'\Rb  \equiv \frac{1}{2\,\bg^2\,\kappa}  \zeta$
  \end{boldmath}
  
  we have after taking the integral over $\Sigma$ using the method of steepest descent:
  \beq \label{MDJ6} 
 \sigma^{\mbox{\tiny AGK}}_n\Lb z',z'_d\Rb  =\frac{C^2 N\Lb z'_d\Rb}{(n-1)!} \Lb \frac{1}{N\Lb z \Rb}\Rb^{3/4} \,  e^{  - \frac{1}{4\,\bg^2\,\kappa}\zeta^2 } \frac{\Lb \frac{1}{2\,\bg^2\,\kappa}\zeta\Rb^{n+1}}{\Gamma\Lb \frac{1}{2\,\bg^2\,\kappa}\zeta \Rb\,}\, \,\exp\Lb -  2 \sqrt{\frac{\frac{1}{2\,\bg^2\,\kappa}\zeta}{N\Lb z'\Rb}}\Rb \eeq

    \begin{boldmath}
II.   For   $N\Lb z' \Rb\,\,>\,n \,>\,\,\frac{1}{2\,\bg^2\,\kappa}\ln N\Lb z'\Rb  \equiv \frac{1}{2\,\bg^2\,\kappa}  \zeta$
  \end{boldmath}

one can see that we have the same equations for the saddle points as have been discussed above (see 
\eq{AGKSPOM1}, \eq{AGKSPDE1}, \eq{AGKSPOMS1} and \eq{AGKSPDES1}). Using these input we obtain for  $ \sigma^{\mbox{\tiny AGK}}_n\Lb z'\Rb $:
    \bea \label{MDJ7} 
 &&\sigma^{\mbox{\tiny AGK}}_n\Lb z, z'_d\Rb  =\\
 &&C^2 N\Lb z'_d\Rb \Lb \frac{n}{N\Lb z \Rb}\Rb^{3/4}\!\!\!\!\!\!\!\!\, \exp\Bigg\{ - \frac{1}{4\,\bg^2\,\kappa}\Lb \zeta + 2 \ln \Lb \frac{\zeta}{2 \,\bg^2\,\kappa}\Rb - \ln n \Rb^2 -\frac{1}{4\,\bg^2\,\kappa}\ln^2 n \Bigg\}\exp\Bigg( -  2 \sqrt{\frac{n}{N\Lb z'\Rb}} \Bigg)\nn
 \eea  

In the region III:
  
    \begin{boldmath}
III.  $n\,>\, N\Lb z' \Rb$
  \end{boldmath}
  
  we have the same situation as in the region II: the solutions for the saddle points of $\Sigma$ and $\Delta$ coincide with  \eq{AGKSPOMS2}   and \label{AGKSPDES2} leading to the following $\sigma^{\mbox{\tiny AGK}}_n\Lb z, z'_d\Rb $:  
  
  \beq \label{MDJ8} 
 \sigma^{\mbox{\tiny AGK}}_n\Lb z, z'_d\Rb  =C^2 N\Lb z'_d\Rb \Lb \frac{n}{N\Lb z \Rb}\Rb^{3/4}\!\!\!\!\exp\Bigg\{ - \frac{1}{4\,\bg^2\,\kappa}\Lb  \zeta^2 \, +  \ln^2 \Lb\frac{ n}{ \frac{1}{\bg^2 \kappa}\ln n} \Rb\Rb\Bigg\} \, \,\exp\Bigg( -  2 \sqrt{\frac{n}{N\Lb z'\Rb}} \Bigg) \eeq

    For finding the multiplicity distributions we need to calculate the total inclusive cross section, viz.:  
    \beq \label{MDJ9}
 \sigma^{\mbox{\tiny AGK}}_{\rm  incl}\Lb z',z'_d\Rb\,\,=\sum^\infty_{n=1} \sigma^{\mbox{\tiny AGK}}_n\Lb z',z'_d\Rb.
 \eeq
    Summing over $n$ \eq{MDJ4} we obtain:
       \beq \label{MDJ10}
 \sigma^{\mbox{\tiny AGK}}_{\rm  incl}\Lb z',z'_d\Rb\,\,= \,\,C^2 N\Lb z \Rb N\Lb z'_d\Rb\eeq 
    
    Therefore, we have the same situation as with \eq{AGK17}: the inclusive cross section has the same form as it follows from the general features of the AGK cutting rules. From this we can also conclude that $C^2=1$ or better to say that we can use the ratio    $\sigma_n/\sigma_{incl}$ in which there is no unknown constants.
    Hence
     \beq \label{MDJ9} 
\mathcal{P}^{J/\Psi}_n= \frac{ \sigma^{\mbox{\tiny AGK}}_n\Lb z, z'_d\Rb}{ \sigma^{\mbox{\tiny AGK}}_{\rm  incl}\Lb z',z'_d\Rb}  = \frac{1}{N\Lb z \Rb}\Lb \frac{n}{N\Lb z \Rb}\Rb^{3/4}\!\!\!\!\exp\Bigg\{ - \frac{1}{4\,\bg^2\,\kappa}\Lb  \zeta^2 \, +  \ln^2 \Lb\frac{ n}{ \frac{1}{\bg^2 \kappa}\ln n} \Rb\Rb\Bigg\} \, \,\exp\Bigg( -  2 \sqrt{\frac{n}{N\Lb z'\Rb}} \Bigg) \eeq      
    One can see that $\mathcal{P}_n$ has the KNO scaling behaviour \cite{KNO}.

    From \eq{MDJ9} and \eq{AGK18} we obtain that 
    \beq \label{CED1}
    \frac{\mathcal{P}^{J/\Psi}_n}{\mathcal{P}^{gluon}_n} = n \frac{n}{N\Lb z \Rb}
    \eeq
    has a simple form.We need to draw the reader's attention to two features of \eq{CED1}.   First it does not reproduce the result of section II where this ratio has been  calculated in the one dimensional model. Second. \eq{CED1} has the same form as it has been discussed in Ref. \cite{GOLEPSI}. Summing two diagrams of \fig{mpsipsi} we reproduce Eq.33 of this reference. Bearing this in mind we refer our reader to this paper for comparison with the available experimental data.
    
\section{Conclusions }
The main result of the paper is \eq{MDJ3}(see \fig{mpsipsi}) which gives the procedure of calculation of the inclusive production using $t$-channel unitarity in high energy QCD (see \eq{MPSI}). We demonstrated this procedure using the dipole densities which have been suggested in our previous publications \cite{LEDIDI,LEDIA,LEAA}. The expressions for these densities  stem from our
attempts to reconcile the exact solution to the Balitsky-Kovchegov (BK) equation\cite{B,K,KOLEB}, which describes
the rare fluctuation \cite{MPSI}  in the dipole-target scattering, with the fact that this equation sums the 'fan'  Pomeron diagrams. We believe that this equation has a  general application for the dipole densities which can be found with more refined procedure. For example we can use the dipole densities given by JIMWLK equation\cite{JIMWLK}, since they can be expressed through the BFKL Pomeron contributions \cite{AKLL,AKLL1}.

We calculate the multiplicity distributions of  produced gluons both with and without $J/\Psi$ inc;usive production.These distributions are defined as the following ratios:
 \begin{subequations}   \bea 
\mbox{For $J/\Psi$ production:} & \mathcal{P}^{J/\Psi}_n = &\frac{ \sigma^{\mbox{\tiny AGK}}_n\Lb z, z'_d\Rb}{ \sigma^{\mbox{\tiny AGK}}_{\rm  incl}\Lb z',z'_d\Rb};\label{CON1}\\
\mbox{Without ~~ $J/\Psi$ :} & \mathcal{P}^{gluon}_n = &\frac{ \sigma^{\mbox{\tiny AGK}}_n\Lb z \Rb}{ \sigma^{\mbox{\tiny AGK}}_{\rm  incl}\Lb z\Rb};\label{CON2}
\eea
\end{subequations}      
The general feature of these distributions that they show the KNO scaling behaviour\cite{KNO}, being the function of $n/N\Lb z \Rb$ where $N\Lb z'\Rb = 2 \exp\Lb \bg z\Rb$ is the average multiplicity of dipoles in the BFKL cascade. 

The ratio of \eq{CED1} has been predicted in almost all previous publications on the subject\cite{KMRS,MOSA,LESI,LSS,GOLEPSI}. Hence we refer our reader to these papers for comparison with the experimental data. We hope that $ \mathcal{P}^{J/\Psi}_n  $  itself can be measured  and in  comparison with these measurements it  will allow us to find the range of energy for our predictions.

In this paper we also found the  multiplicity distribution in the simple one dimensional model which corresponds to the BFKL cascade in QCD.  The main question that we answered is how the production of many Pomeron ladders in \fig{nlad} contribute to the production of $J/\Psi$ at fixed $n$. They all cancel after summing over $n$. Our result is that the influence of  final $n$ is not very important and they can be treated as corrections.

We hope that this paper will contribute to the further study of the Pomeron calculus in QCD, especially to the understanding of  typical scales for tinclusive production. In this paper we find the inclusive cross section integrated over transverse momenta. We are planning to study the production of the gluons and quarkonias with fixed transverse momenta ($p_T$) to learn more about suppression at $p_T < Q_s$ and about violation of $k_T$-factorization\cite{CCH,COEL,LRSS}.

{\bf  Acknowledgements}
 
We thank our colleagues at Tel Aviv university for discussions. Special thanks go A. Kovner and M. Lublinsky
for stimulating and encouraging discussions on the subject of this paper. This research was supported by BSF grant
2022132

\end{document}